\documentclass{ws-procs9x6}

\begin{document}

\title{Nuclear Data Resources for Capture gamma-Ray Spectroscopy and Related Topics}

\author{B. Pritychenko$^*$}

\address{National Nuclear Data Center, Brookhaven National Laboratory \\ 
Upton, NY 11973-5000, USA \\
$^*$E-mail: pritychenko@bnl.gov}

\begin{abstract}
Nuclear reaction data play an important role in nuclear reactor, medical, and fundamental science and 
national security applications. The wealth of information is stored in internally adopted ENDF-6 
and EXFOR formats. We present a complete calculation of resonance integrals, Westcott factors, thermal 
and Maxwellian-averaged cross sections for Z=1-100 using evaluated nuclear reaction data. The addition of 
newly-evaluated neutron reaction libraries, and improvements in data processing 
techniques allows us to calculate nuclear industry and astrophysics parameters, and provide additional 
insights on all currently available neutron-induced reaction data. Nuclear reaction calculations will be discussed 
and an overview of the latest reaction data developments will be given.
\end{abstract}


\bodymatter

\section{Introduction}

The value of compilation, evaluation and computer storage of neutron cross section data was first recognized in the 
early 50's \cite{70Pear} prompted by the urgent needs of nuclear industry. These cross sections  were summarized in BNL-325 report 
and Evaluated Nuclear Data File (ENDF) library \cite{06Mugh,06Chad}. Over the  years, the nuclear data 
activities were extended to all low- and intermediate-energy nuclear physics topics. 

Computer storage  and worldwide dissemination of nuclear data strongly depend on the presently available computer technologies \cite{06Pri} 
and data format developments. In present days, size and representation of nuclear data files are no longer limited by the computer hardware and software. 
However, for historic and consistency reasons, nuclear data are still stored in 80-character long formats.

For years, nuclear data research and developments were driven by nuclear science and technology applications. These developments evolved and lead to 
the creation of complementary products such as Experimental Nuclear Reaction Data (EXFOR) database \cite{EXFOR}. The overall 
nuclear data improvements, maturation  and modern computer technologies created many new opportunities for nuclear reaction calculations, data mining 
and analysis. For the capture $\gamma$-ray spectroscopy purposes, we will concentrate on the selected nuclear astrophysics and reactor operation 
integral quantities that can be extracted from the beta version of ENDF/B-VII.1$\beta$3 library. Current results will 
be finalized by the end of this year during the public release of ENDF/B-VII.1 library \cite{11Chad}.     
\begin{figure}
\centering
\includegraphics[width=0.8\columnwidth]{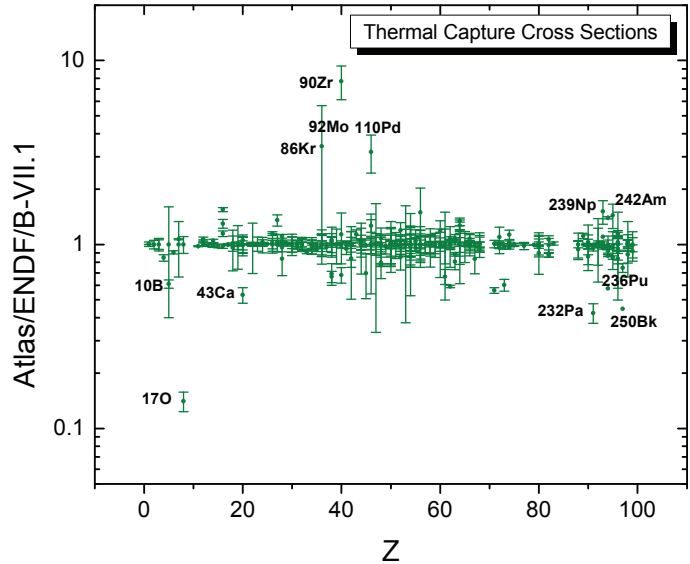}
\caption{Ratio of Atlas of Neutron Resonances \cite{06Mugh} and ENDF/B-VII.1$\beta$3 thermal neutron capture cross sections.}
\label{fig1}
\end{figure}

\section{ENDF Integral Quantities}

Neutron capture cross sections govern the production of chemical elements in the AGB and red giant stars, 
safe operation of nuclear reactors, serve in nuclear structure measurements, etc.   The low-energy neutron cross section 
values are often influenced by the contributions from resolved and unresolved resonance regions. 
To estimate these contributions across the whole ENDF/B-VII.1 library range of nuclei \cite{11Chad} and provide 
additional insights on the data quality for nuclear reactor and astrophysics applications, we 
have selected thermal and Maxwellian-averaged cross sections, resonance integrals and Westcott factors \cite{06Mugh,10Pri} for 
calculation and further analysis. ENDF/B-VII.0 \cite{06Chad} and ENDF/B-VII.1$\beta$3 evaluated neutron cross sections were Doppler 
broadened using the code PREPRO \cite{07Cul} with the precision of 0.1\%.  These reconstructed and linearized data were used to 
calculate the selected quantities using the definite Java integration method \cite{10BPr}.

The ratio of thermal $\sigma$(n,$\gamma$)  is shown in Fig. \ref{fig1}. 
Using the method of visual inspection we notice the deviations for light and medium nuclei and minor actinides  evaluations. These differences, in the low- and medium-Z region, are 
attributed to the lack of or insufficient experimental data for $^{10}$B, $^{17}$O, $^{43}$Ca, $^{86}$Kr, $^{110}$Pd, deficiencies for $^{58}$Co,$^{92}$Mo  and recent  re-evaluation of $^{90}$Zr.   
While, in the actinide region, deviations are due to new evaluations from the Actinoid file \cite{08Iwa}.

\begin{figure}
\centering
\includegraphics[width=0.8\columnwidth]{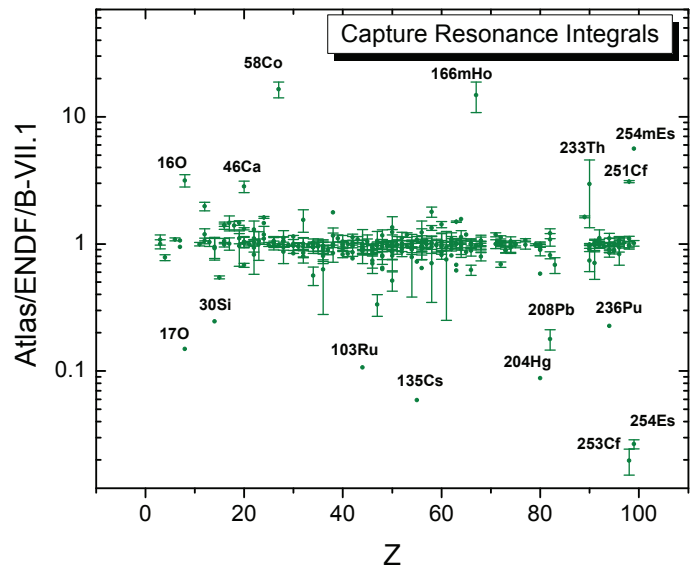}
\caption{Ratio of Atlas of Neutron Resonances \cite{06Mugh} and ENDF/B-VII.1$\beta$3 thermal neutron capture resonance integrals.}
\label{IQ.3}
\end{figure}

Neutron capture resonance integrals were calculated for 0.5 eV - 20 MeV incident neutron energy range and shown in Fig. \ref{IQ.3}. 
Several data outliers in this case could be traced to the lack of measurements and incomplete overlap of experimental and theoretical data  for $^{17}$O, $^{103}$Ru ,$^{166m}$Ho and $^{46}$Ca, $^{58}$Co, $^{135}$Cs, $^{204}$Hg, respectively. 
However, there are neutron capture cross section deficiencies in the keV region of energies for $^{16}$O,$^{30}$Si and $^{208}$Pb evaluations.

\begin{figure}
\centering
\includegraphics[width=0.8\columnwidth]{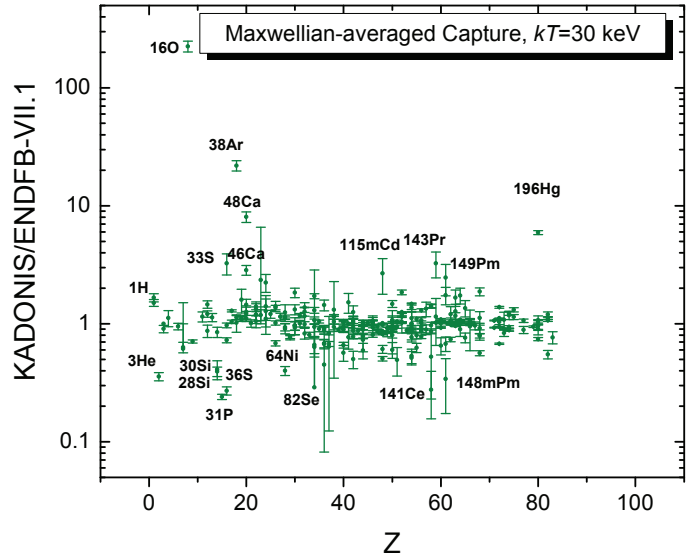}
\caption{Ratio of Karlsruhe Astrophysical Database of Nucleosynthesis in Stars (KADONIS) \cite{06Dil} and ENDF/B-VII.1$\beta$3 Maxwellian-averaged cross sections at {\it kT}=30 keV.}
\label{IQ.5}
\end{figure}

Maxwellian-averaged cross sections play an important role in power reactor developments and slow-neutron capture ({\it s}-process) nucleosynthesis calculations \cite{00Bao}. 
The $s$-process is mostly responsible for element formation in stars from $^{56}$Fe to $^{209}$Bi.  The detailed analysis of the Fig. \ref{IQ.5}
 data \cite{10Pri} demonstrates the nuclear astrophysics potential of ENDF libraries as a complimentary source of evaluated cross sections and reaction rates. 
There are noticeable differences between KADONIS \cite{06Dil} and ENDF/B-VII.1$\beta$3 libraries for light  and medium nuclei.  
$^{1}$H deviation is due to differences between center of mass and lab system cross section values. For $^{28,30}$Si, $^{31}$P, $^{64}$Ni and $^{196}$Hg KADONIS values are based on a single recent measurement. 
 Due to lack of experimental data theoretical values were adopted for $^{38}$Ar, $^{82}$Se, $^{115m}$Cd, $^{141}$Ce, $^{143}$Pr and $^{148m,149}$Pm. 
Deficiencies in $^{16}$O, $^{46,48}$Ca and $^{33,36}$S  originate from the old or insignificant for integral tests ENDF evaluations and coverage problems in EXFOR database \cite{EXFOR}, respectively. 

\begin{figure}
\centering
\includegraphics[width=0.8\columnwidth]{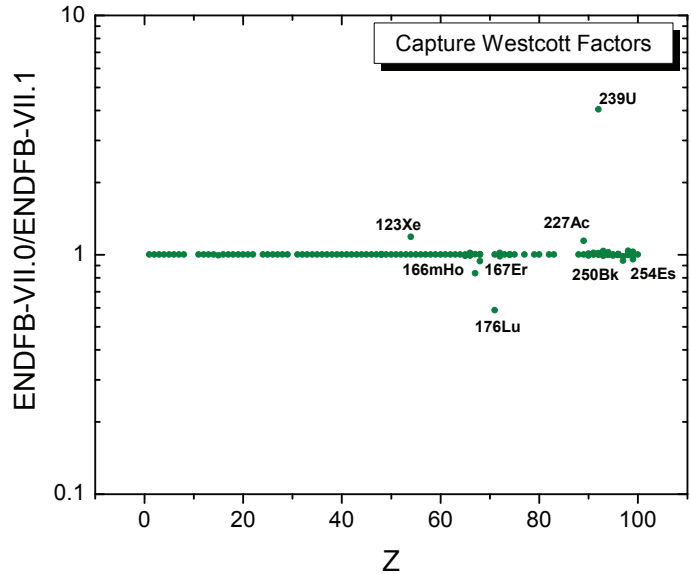}
\caption{Ratio of thermal neutron capture Westcott factors between ENDF/B-VII.0 and ENDF/B-VII.1$\beta$3 libraries.}
\label{IQ.6}
\end{figure}

Shown in Fig. \ref{IQ.6} ratio of capture Westcott factors indicate large deviations for $^{239}$U and $^{176}$Lu. 
These  deviations reflect the changes in ENDF/B-VII.1$\beta$3 library where Westcott factors evolved from 3.997 to 0.989 and 
from 1.002 to 1.711 for $^{239}$U and $^{176}$Lu, respectively. The last number agrees well with the recommended value of 1.75 \cite{06Mugh}. 
Smaller deviations as in $^{123}$Xe are due to adoption of new evaluations in ENDF/B-VII.1$\beta$3 library and lack of experimental data for this nucleus.

\section{Conclusion \& Outlook}
The present work demonstrates large potential of ENDF libraries for capture $\gamma$-ray spectroscopy applications. Maxwellian-averaged and thermal 
capture cross sections, resonance integrals and Westcott factors have been extracted from the beta version of ENDF/B-VII.1$\beta$3 library. 
These results are important in fundamental science, nuclear technologies and  for nuclear data validation. 
Several $\beta$-version $\sigma$(n,$\gamma$) deficiencies ($^{16}$O, $^{58}$Co, ...) will be resolved during the official library release. 
The new ENDF/B-VII.1 library \cite{11Chad} will be publicly available in December of 2011 at the National Nuclear Data Center website {\it http://www.nndc.bnl.gov} \cite{06Pri}. Future nuclear reaction data mining work  
will involve extensive analysis of neutron cross section covariance files and extended coverage for neutron capture, fission and elastic 
scattering and all major neutron libraries.

\section*{Acknowledgments}
The author is grateful to M. Herman, S.F. Mughabghab and M. Blennau (BNL) for support of this work, 
productive discussions and careful reading of the manuscript and useful suggestions, respectively. 
This work is supported by the Office of Nuclear Physics, Office of Science of the U.S. Department of Energy, 
under contract no. DE-AC02-98CH10886 with Brookhaven Science Associates, LLC.

\end{document}